\documentclass{IEEEcsmag}
\usepackage[colorlinks,urlcolor=blue,linkcolor=blue,citecolor=blue,breaklinks=true]{hyperref}

\usepackage{upmath}
\usepackage{booktabs}

\jvol{XX}
\jnum{XX}

\newcommand{\RNum}[1]{\uppercase\expandafter{\romannumeral #1\relax}}

\begin{document}

\title{Investigative Pattern Detection Framework for Counterterrorism}


\author{Shashika R. Muramudalige}
\affil{Department of Electrical \& Computer Engineering, \\ Colorado State University}

\author{Benjamin W. K. Hung}
\affil{Department of Electrical \& Computer Engineering, \\ Colorado State University}

\author{Rosanne Libretti}
\affil{Western Jihadism Project, \\ Brandeis University}

\author{Jytte Klausen}
\affil{Western Jihadism Project, \\ Brandeis University}

\author{Anura P. Jayasumana}
\affil{Department of Electrical \& Computer Engineering, \\ Colorado State University}


\begin{abstract}

Law-enforcement investigations aimed at preventing attacks by violent extremists have become increasingly important for public safety. The problem is exacerbated by the massive data volumes that need to be scanned to identify complex behaviors of extremists and groups. Automated tools are required to extract information to respond queries from analysts, continually scan new information, integrate them with past events, and then alert about emerging threats. We address challenges in investigative pattern detection and develop an Investigative Pattern Detection Framework for Counterterrorism (INSPECT). The framework integrates numerous computing tools that include machine learning techniques to identify behavioral indicators and graph pattern matching techniques to detect risk profiles/groups. INSPECT also automates multiple tasks for large-scale mining of detailed forensic biographies, forming knowledge networks, and querying for behavioral indicators and radicalization trajectories. INSPECT targets human-in-the-loop mode of investigative search and has been validated and evaluated using an evolving dataset on domestic jihadism.

\end{abstract}

\maketitle

\chapterinitial{Counterterrorism} professionals, law enforcement authorities, and behavioral research scientists are hampered by the lack of efficient and scalable computing tools to mine and process relevant information and identify violent extremists and emerging threats. Due to their dynamic and complex behavioral patterns, analysis, modeling, detection, and prediction of relevant events and profiles involve scanning, filtering, cataloging, and mining massive amounts of data.
For authorities, the major challenge is to identify emerging terrorist threats based on surveillance data and behavioral events covering millions of people. The robust and accurate detection is an imperative to preserve public trust and acceptance. In 2019, the then acting director of the National Counterterrorism Center described the problem, ``my ops [operations] center receives something in excess of 10,000 terrorism-related intelligence reports a day through which we need to sift. And those 10,000 reports contain 16,000 names. Daily.''~\cite{travers_washington_inst}. New information often makes events that were considered innocuous months or years ago relevant to investigations. Human intuition cannot compute massive volumes of data on the scale  described.

Threat risk assessments by law enforcement analysts are centered on understanding individuals on five critical variables: intent, history, capability, opportunity, and resolve. Analysts are responsible
for conducting assessments with respect to each of these variables, knowing when changes in any one occur, and if the changes signal
whether an individual is at more or less risk of conducting a terrorist attack. Thus both discovery and knowledge management are critical factors in supporting analysts in these assessments and for ultimately enabling law enforcement to prevent future attacks.

However, analysts face significant challenges with respect to both discovery and knowledge management~\cite{understanding_rel_alqaeda_isss_subrahmanian}. The first is simply scalability--how to produce salient risk assessments that are derived from the sheer volume of disparate types
of data~\cite{protect_homeland:1}. The data available are often found in disconnected portals, consisting of uploaded text reports from the field and not structured such that they can be fused, normalized, and co-referenced to enable the analysis needed for these risk assessments. The other significant challenge is dynamicity--the dynamical nature of the threat as well as individual behavioral indicators necessitate databases to be updated very frequently~\cite{threat_assessment}, which in turn requires more automated analysis capabilities, to include scaling and speeding up structured information extraction related to radicalization indicators, as well as fusing knowledge such that it is relatable and contextualized.


We introduce a novel \textbf{Investigative Pattern Detection Framework for Counterterrorism (INSPECT)} to aid  human analysts sort signal from noise in detecting groups and individuals with  behavioral patterns of interest. The framework consists of several computing tools for automating the investigative pattern detection through machine learning and graph pattern matching techniques. The value of the combination of a machine learning approach to text mining fused with a validated sociological model of radicalization and with network science is the ability of computational methods to detect behaviorally meaningful signals from large amounts of data at great speed. The sociological data used to profile extremist behaviors used here derives from multi-year study of domestic jihadists by one of the authors.

The proposed techniques and the framework may be adapted to other investigative detection contexts with appropriate datasets. Collaboration between social scientists and computer scientists is critical to the effort to build evidence-based machine learning
models of complex human behavior. In such collaborative research, human-in-the-loop (HITL) plays a key role that allows domain expertise
(social scientists) to validate the reliability and accuracy of the proposed investigative pattern detection framework by computer scientists. Further, HITL is an iterative process to improve the
capabilities of the framework while sharing the knowledge and the experience between social and computational contexts.

\begin{figure*}
\centerline{\includegraphics[width=37pc]{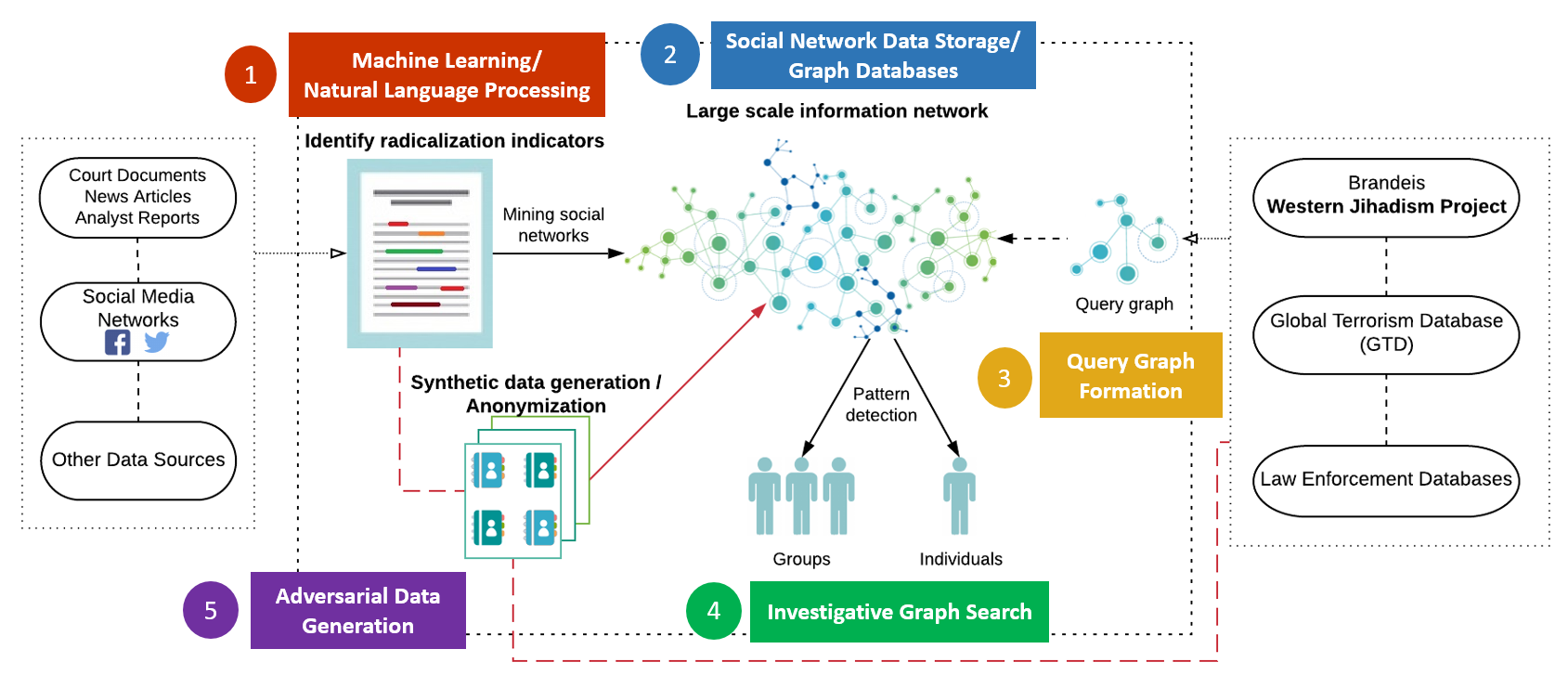}}
\caption{High-level architecture of INSPECT (Investigative Pattern Detection for Counterterrorism).}\label{fig:framework}
\end{figure*}

The use of behavioral markers to profile extremist radicalization has been subject to controversy on privacy ground but is now widely
accepted by social scientists and law enforcement [9]. (Radicalization refers to the process by which people come to support violent extremism and join terrorist groups or commit a terrorist act.)
Guidelines issued by the Office of the Director of National Intelligence with collaboration by the FBI, the US National Counterterrorism Center, and the US Department of Homeland Security,
enumerates a list of observable indicators of potential violent extremists drawn from research. A study released by the FBI in 2019 of 52 ideologically-motivated lone offenders concluded that “they traveled down the same observable and discernable pathways to violence as other attackers”~\cite{lone_offender_fbi}. The behavioral model of radicalization used here was developed in previous research by the authors~\cite{rad_pathways_prob:1}.To be clear, we are not dealing with metadata used for the purpose of mass surveillance. Ours is an effort to use advances in machine learning, natural language processing (NLP), and graph data bases to extract risk profiles from an evidence-based sociological model of human behaviors known to be associated with escalating risk of violent action~\cite{dynamic_threat_assessment_klausen:1}.

\section{INSPECT ARCHITECTURE}


To address the critical needs of scanning and mining large volumes of data while adapting to scale and dynamicity and to support human-in-the-loop investigative searches, we develop an end-to-end investigative pattern detection framework INSPECT (Investigative Pattern Detection Framework for Counterterrorism). Natural Language Processing (NLP) techniques are integrated and adapted in INSPECT to extract radicalization behavioral indicators and features (such as timestamps, relationships) in different text sources. The extracted data are arranged to a knowledge network consisting of behavioral patterns of individuals of interest. Machine learning and graph pattern matching techniques are used in INSPECT for investigative graph search to identify individuals and groups from large information and knowledge networks. The solution is built on top of a graph database, with its convenient data storing mechanism, to scalability for dynamically manipulating large, heterogeneous networks. Further, we introduce a novel synthetic profile generation technique for behavioral patterns to address two intrinsic challenges in the domain: (I) the lack of sufficient training data for both human coders (whose expertise pertain to identification and classification of indicators in documents) and machine learning models, and (II) the need for shareable anonymized data sets of different sizes and characteristics that do not violate various privacy regulations or constraints.

The overall architecture and components of the INSPECT  framework is illustrated in Figure~\ref{fig:framework}. It consists of  the following major functional components:  

\begin{enumerate}
    \item \textit{NLP to identify radicalization indicators in text sources:} The details of extremists are mostly available as news articles, court documents, and reports, which are disparate text sources. The data consists of various behavioral indicators in different stages of radicalization. We apply several NLP techniques to extract the behavioral indicators and other information. 
    \item \textit{Graph databases:} The extracted data from text sources are innately captured in the form of a social network that consist individuals together with their behavioral indicators, as well as links connecting individuals,  organizations and behavioral indicators. This dataset is highly linked and storing and processing such data is a challenging task. Graph databases, designed for pattern-based querying over huge volumes, contains many features for mining such networks. Therefore, we use a graph database over SQL and NoSQL databases where considers the relationships (links) between data as equally important as the data itself.
    \item \textit{Query graph formation:} With years of experience observing the extremists’ behavior, social scientists have studied the diverse patterns of radicalization. We use their empirical knowledge to model query graphs representing the ordinary or specific behavior of an extremist. The example used here draws in the relational ontology developed by one of the authors assess contagion networks in a jihadist population.
    \item \textit{Investigative graph search:} We have developed and implemented a set of algorithms to explore potentially risky individuals and groups on knowledge networks~\cite{enhancing_pings:1}. Graph searches are performed as custom queries to the graph databases, which enhances the efficiency and the scalability of data processing while utilizing the database features.
    \item \textit{Adversarial data generation:} We propose a novel synthetic data generation technique to mimic the behavior of extremists. This method is widely applicable in other domains as well where the available datasets are small, sparse, or insufficient.
\end{enumerate}

\begin{figure*}
\centerline{\includegraphics[width=34pc]{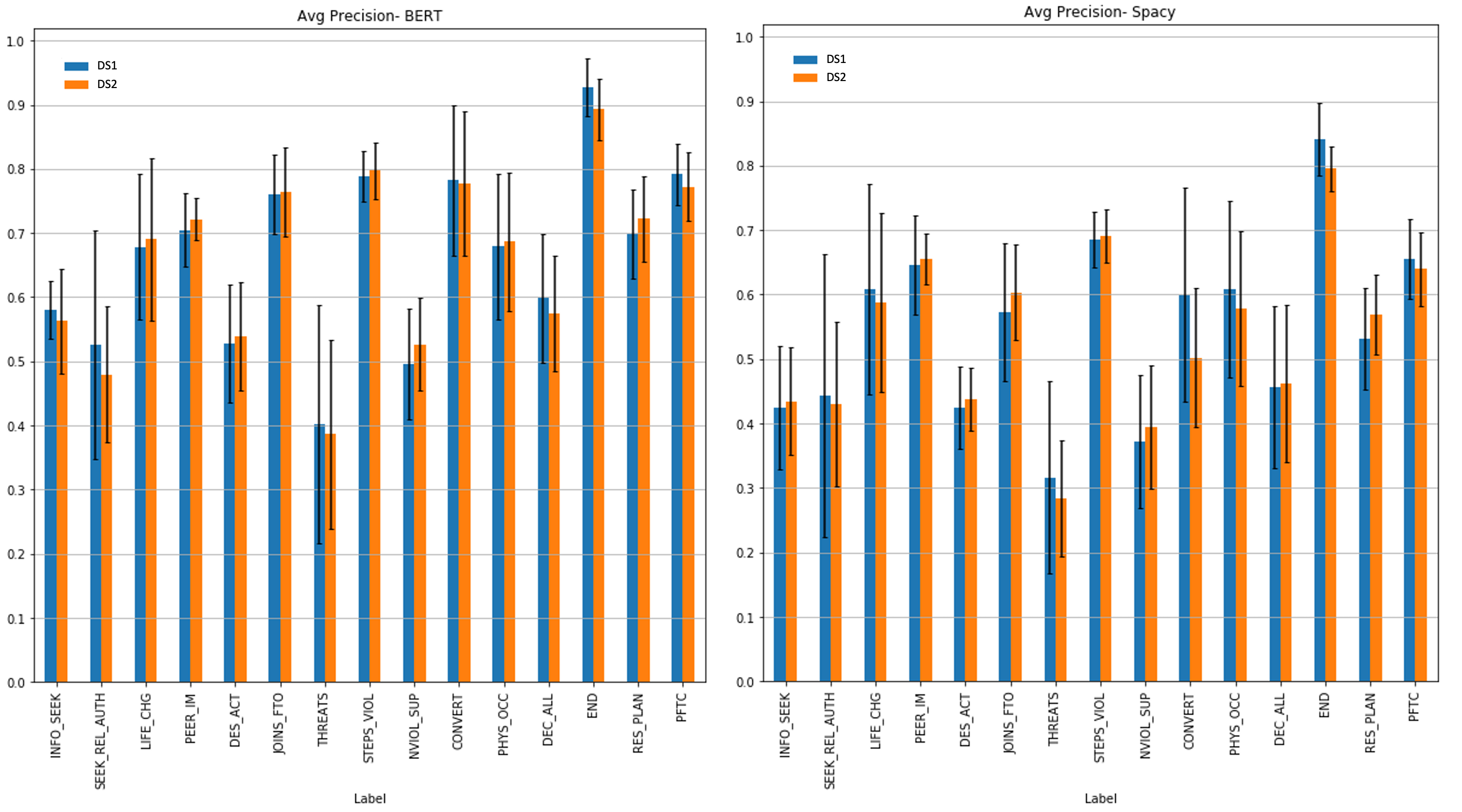}}
\caption{Each indicators' classification precision across 10-folds for both \texttt{BERT} and \texttt{SpaCy}. x-axis represents 15 different behavioral indicators (labels). Two training datasets are shown; DS1 (blue) and DS2 (orange).}\label{fig:AveragePrecisions}
\end{figure*}

\subsection{Extraction of behavioral indicators}
Investigating the behaviors of already-identified violent extremists is a crucial instrument for recognizing emerging  profiles based on similar behavioral patterns. Most such data is available in public domain text sources, which may be mined to study their behavioral nature without violating confidentiality restrictions. Social scientists rely  on official press releases, court documents, trusted news sources and verified social media accounts of extremists. The common practice is for researchers to read and inspect related text documents and then manually label the behavioral indicators present (a process known as ‘coding’). The training dataset  and text cues used for the research and to train NLP algorithms  was provided by the social scientists at the Western Jihadism Project (WJP), and contains phrases, sentences or short paragraphs which were manually labeled with up to 3 radicalization indicators.    

Among different text classification techniques, we have made significant improvements using multi-label text classification~\cite{hst:1}, a supervised learning model where text documents are assigned one or more trained categories/labels. We focused initially on a sentence-level classification for simplicity and a finer analysis granularity for social scientists. In this work, 15 radicalization indicators are trained on a deep-classification model where returns probability scores for all labels (indicators) in each sentence in the document. We use \texttt{SpaCy} library (a convolutional neural network for NLP) and \texttt{BERT} (a pre-trained transfer learning model for different NLP tasks).

Due to the unbalanced nature of the training data (where some labels are more prevalent than others), a stratified train-test splitting was implemented to ensure that the resulting test set has a proportional set of labeled data akin to the training set as well as 10-fold cross validation.

Figure~\ref{fig:AveragePrecisions} depicts the precision scores with 10-folds for each of the labels for both the \texttt{BERT} and \texttt{SpaCy} models using two recent training data sets. Also displayed are the inter-label standard deviations (as error bars) and the intra-label standard deviations (shown below the plots). In both datasets, \texttt{BERT} outperforms \texttt{SpaCy} because of \texttt{BERT}’s ability to distinguish underlying data classes as well as a pre-trained model on a diverse range of generic corpora, even though our datasets are inherently unbalanced and small. Furthermore, the classification results were sent to social scientists for validation as a part of the HITL process. Based on the model results, social scientists fine-tuned the training set by understanding the issues in the dataset. Human validation is essential in the law enforcement domain to produce accurate results with sensitive behavioral data. Therefore, we have gone through multiple iterations to fine-tuned both the training dataset and the NLP models in the HITL process. 

Our implementation also integrates the Named Entity Recognition (NER), an information extraction technique to detect pre-defined entities in text sources as person names, organizations, locations, time, etc. We use the SpaCy NER module
to detect date, time, person names, and organizations. In this case, NER enhances  the capability to find relevant metadata of extremists in diverse text sources, which in turn are embedded while forming knowledge networks. Coreference resolution is another NLP task to find all expressions that relate to the same entity, which we use to extract the similar expressions related to an individual. However, our investigations revealed that  the dynamicity and different lengths of the text sources led erroneous results. We were able to  improve the results using different text preprocessing techniques such as lower casing, stop words removal, lemmatization.  With these NLP tasks, we extracted the relevant information, including behavioral indicators and their relevant metadata, to produce structured knowledge networks. An important observation is the fact that in the  HITL context, these  NLP outcomes significantly help social scientists improve their manual coding process. These enhancements improve their efficiency allowing them to process  considerably higher volumes of text sources while magnifying the knowledge of trajectories, leading to a comparatively larger dataset to train more robust NLP models.

\begin{figure*}
\centerline{\includegraphics[width=37pc]{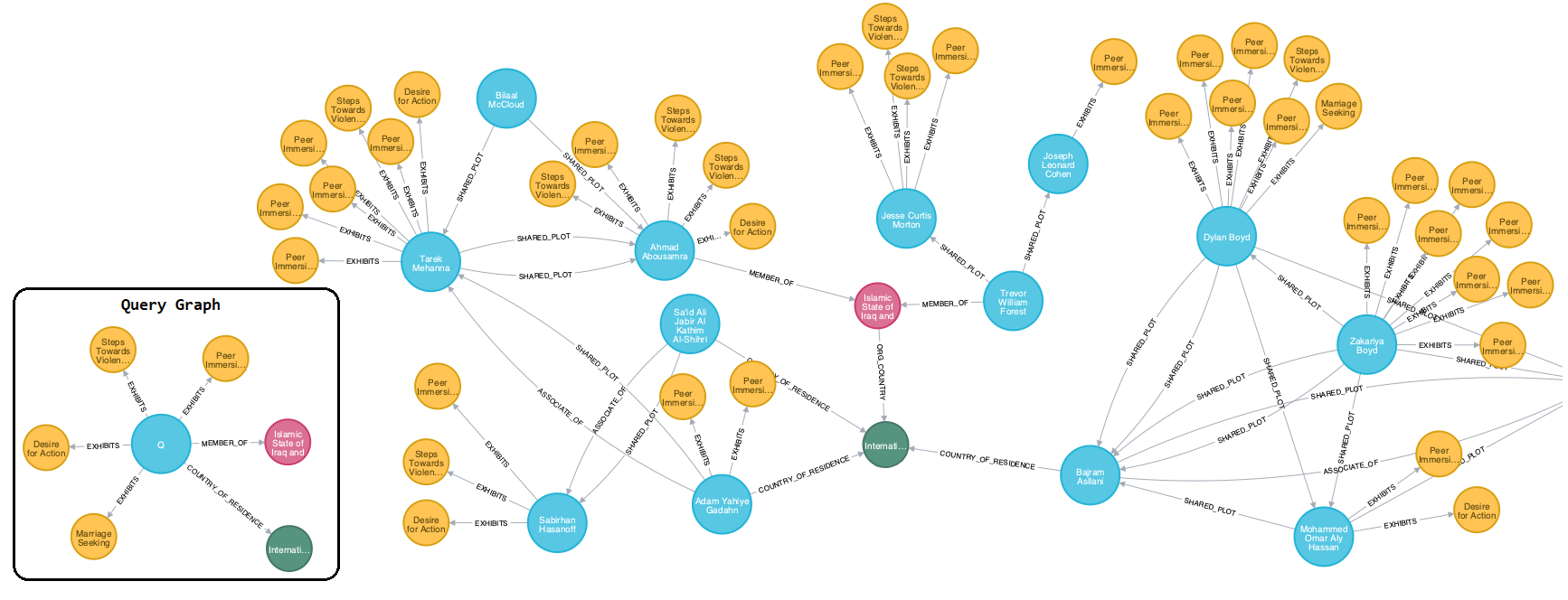}}
\caption{A subset of results for an inexact match \textit{(similarity threshold = 0.7)} neighborhood measure in the WJP graph database based on a given query graph. The highlighted box represents an example query graph. The blue node represents an individual that presents in the network. Yellow nodes represent the behavioral indicators with their types. Green and red nodes depict a country and a terrorist organization, respectively.}\label{fig:inexact_results}
\end{figure*} 

\subsection{Investigative graph search of knowledge networks}
Identification of structured information of behavioral indicators via NLP tools and techniques allows for the formation of an enriched knowledge networks of individuals while dealing with the dynamicity and scalability of behavioral patterns. Moreover, persistent storage necessitates storing such time-critical networks for a long term. We use a \texttt{Neo4j} graph database to store knowledge graphs while adapting to the dynamicity and scalability of the behavioral data. Neo4j is currently the foremost graph database platform and contains various useful features to implement custom graph searches in a specific domain while enabling the consumption of basic graph algorithms~\cite{enhancing_pings:1}. 

An investigative search to detect potential extremists or groups in knowledge networks in the \texttt{Neo4j} graph database complements the pattern detection and analysis. INSPECT relies on  a similarity measure based inexact graph pattern matching technique that we proposed in~\cite{enhancing_pings:1} to investigative individuals and groups. The method addresses the inexact sub-graph isomorphism problem because the similarity measure executes w.r.t a given query graph. The query graphs used in these experiments are based on the relational ontology developed by one of the authors assess contagion networks in jihadist populations.

\begin{figure*}
\centerline{\includegraphics[width=37pc]{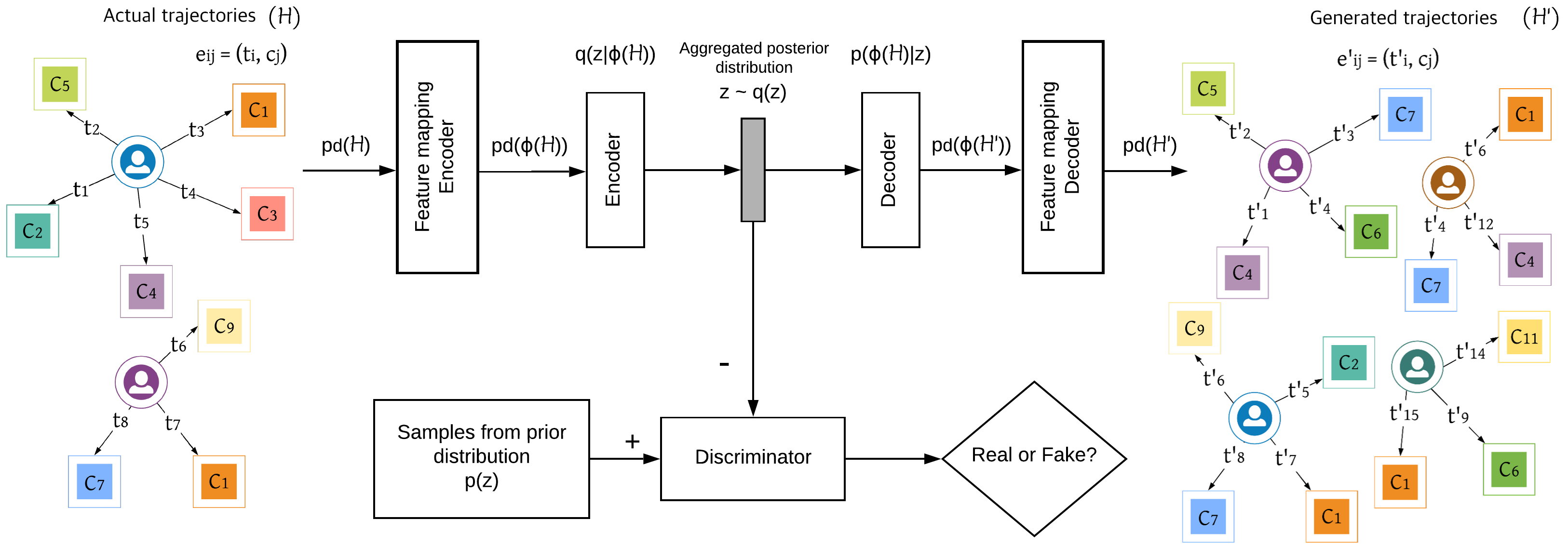}}
\caption{Synthetic data generator architecture for INSPECT  consisting of an ordinary adversarial autoencoder couples with the proposed feature mapping encoder and decoder. The top row depicts the autoencoder that reconstructs the feature-mapped data from the latent code $z$. The second row shows the discriminative network that predicts whether the samples emerge from the hidden code of the autoencoder $q(z)$ or the user-defined prior distribution $p(z)$~\cite{aae:1}. $p_d(\mathcal{H})$ and $p_d(\phi(\mathcal{H}))$ denotes the actual and the feature-mapped data distributions, respectively. $p_d(\phi(\mathcal{H'}))$ denotes the generated feature-mapped data distribution. $p_d(\mathcal{H'})$ denotes the generated data distribution after send through the feature-mapped decoder. $q(z|\phi(\mathcal{H}))$ and $p(\phi(\mathcal{H})|z)$ denote the encoding and decoding distributions of the autoencoder respectively.}\label{fig:aae}
\end{figure*}

An investigative search retrieves a set of suspicious individuals or groups. To this end, our system integrates a vectorized, similarity-based solution approach called INSiGHT~\cite{insight:1} for investigative graph search. With the understanding of the reliability and scalability, we then developed more advanced investigative graph search library that runs on top of the \texttt{Neo4j} database with custom procedures was developed: \textit{individual similarity} (for individuals) and \textit{neighborhood similarity} (for groups)~\cite{enhancing_pings:1}. We further discuss the extensive details of the investigative graph search, including the scalability of this approach with database features.

The identification of suspicious groups is crucial in law enforcement domains. Radicalized extremists are known to conspire  with individuals or groups with similar interests for specific objectives. Searching for single-person subgraphs is therefore not  sufficient to detect a more complete risk assessment that includes collectively suspicious behaviors. Investigators believe, e.g., that the plot in San Bernardino Terrorist Attack in  2015 was not detected before the attack because the perpetrators were radicalized separately before meeting each other. When detecting multiple individuals as a group, their collective behavior could have a magnifying effect in  sudden extremist development or a threat of an attack. To identify such plots, it is vital to detect associations and collective behaviors via the proposed neighborhood measure. Figure~\ref{fig:inexact_results} depicts a subset of the inexact neighborhood measure results in the WJP graph database based on the query graph shown in the highlighted box that represents an example query graph with four different indicators (orange) and two filtering options by the country (green) and the organization (red).
These graph search techniques allow detectives and relevant authorities to focus on a smaller subset of individuals who have a higher likelihood of being extremists, rather than surveilling an enormous number of people, allow them to take necessary actions efficiently within a smaller time frame to avoid such fatal attacks.

\subsection{Synthetic profile generation}
Having only a very limited number of historical profiles of radicalized individuals, and thus a limited number of trajectories available  are major hindrances in the evaluations and studies related to radicalization detection.  Even though many  deep-learning techniques may be applicable to help  identify latent trends in behavioral patterns, such machine learning models still require large datasets for robust and realizable outcomes. Having smaller number of data limits the knowledge of human  coders who manually annotate the text documents. A critically important requirement therefore is a  data generation technique that generates realistic datasets starting with  small and discrete datasets. , which is beneficial in various domains. Further,  synthetically generated  trajectories provide a significant degree of anonymization with respect to original data, and therefore provides an important source of shareable and publishable  behavioral data  without violating any  privacy regulations. Thus synthetic data generation described below  is an integrated part of INSPECT.

We propose an adversarial approach that augments adversarial autoencoder (AAE)~\cite{aae:1} with novel feature mapping techniques. The architecture of our proposed technique is depicted in Figure 4. We consider a behavioral indicator ($e_{ij}$) is consist of a timestamp ($t_i$) and the category ($c_j$), which is the common in radicalization data. Irrespective of the data is inherently small and spare, the proposed feature mapping techniques allows to mitigate such challenges fluently. Feature mapping techniques utilize a cumulative distribution of each category to map the data to a consistent data space. 

We have already demonstrated the effectiveness of the data generation technique for enhancing machine learning tools for other domains  characterized by very limited  data availability and/or  inconsistencies in data. Synthetic data so generated have led to significant improvements in  phishing websites detection, and video traffic classification. Therefore, the proposed approach is widely applicable. In radicalization detection, synthetic data generation is extremely useful to improve the algorithms, and also to train computer and social scientists on different aspects of radicalization.

With all these components in the proposed investigative pattern detection framework, we have also verified  the framework's robustness and efficiency over diverse datasets. Some of the library routines are publicly available in \url{https://github.com/cnrl-csu}.  We have demonstrated that the NLP techniques, the storage selection, and investigative graph search were implemented with the consideration of reliability and scalability.  Further, the proposed modularized architecture of INSPECT  (Fig.~\ref{fig:framework}) maintains the generalizability of the detection process that allows us  to add/update/disable different computing components based on the data and the requirements. Therefore, the proposed framework is capable of grasping diverse datasets and adapting any new or future technologies.

\section{CONCLUSION}

We presented INSPECT, a set of computation tools integrated as a framework for investigative pattern detection aimed at flagging  suspicious individual and group profiles. INSPECT automates behavioral pattern detection by using diverse machine learning and graph pattern matching based techniques for tasks such as  data extraction in text sources, data storage, query graph formation, investigative graph search, and synthetic profile generation. We further explained an approach to overcome the challenges with small and sparse datasets, which are inherent in behavioral pattern data. The  proposed framework has been validated using detailed biographies of known domestic jihadists in the Western Jihadism Project (WJP) database.  

Ongoing and future work include applying the framework to a wide range of other investigative domains while maintaining accuracy and scalability. We also plan to  implement a web application to embed the framework with user-friendly interfaces that will be significantly beneficial to social scientists and investigators.

\section{ACKNOWLEDGMENT}

This work was supported by the U.S. Department of Justice, Office of Justice Programs/National Institute of Justice under Award 2017-ZA-CX-0002. Opinions or points of view expressed in this article are those of the authors and do not necessarily reflect the official position of policies of the U.S. Department of Justice.

\bibliographystyle{IEEEtranS}
\bibliography{inv_pattern}

\vspace{3em}

\begin{IEEEbiography}{Shashika R. Muramudalige}{\,} graduated with his Ph.D. in computer engineering in the spring of 2022 from Department of Electrical and Computer Engineering, Colorado State University, USA. He received the M.Sc. degree in Computer Science and Engineering from University of Moratuwa, Sri Lanka in 2018. His research interests include social network analysis, data science, and machine learning. Contact him at shashika.muramudalige@colostate.edu.
\end{IEEEbiography}

\begin{IEEEbiography}{Benjamin W. K. Hung}{\,}is a Post-Doctoral Fellow with Colorado State University and an Operations Research/Systems Analysis Officer with the United States Army, Washington, DC, USA. He received the Ph.D. degree in Systems Engineering from Colorado State University, USA. His current research interests include artificial intelligence, pattern detection in social systems, social network analysis, and computational social systems. Contact him at benjamin.hung@colostate.edu.
\end{IEEEbiography}

\begin{IEEEbiography}{Rosanne Libretti}{\,} is a research specialist at the Western Jihadism Project at Brandeis University. She graduated from the John Jay College of Criminal Justice (CUNY) with a M.A. in forensic psychology. She has worked as a research assistant on projects covering a wide range of topics including sex offender recidivism, Jihadist radicalization trajectories, law enforcement officer personality, and homicides involving sex worker victims.
Contact her at roselib@brandeis.edu.
\end{IEEEbiography}

\begin{IEEEbiography}{Jytte Klausen}{\,} is the Lawrence A. Wien Professor of International Cooperation at Brandeis University and an Affiliate at the Center for European Studies at Harvard University. She earned her Ph.D.s at the New School for Social Research in New York, USA and University of Aarhus, Denmark. In 2006, she founded the Western Jihadism Project, which studies Western violent extremists associated with al-Qa'ida. she has written for Foreign Affairs, the Wall Street Journal, the Boston Globe, and other national and international outlets, and she is a regular commentator on the BBC, Voice of America, and other U.S. and international media.
Contact her at klausen@brandeis.edu.
\end{IEEEbiography}

\begin{IEEEbiography}{Anura P. Jayasumana}{\,} is a Professor in Electrial \& Computer Engineering at Colorado State University where he holds joint appointments in Computer Science and Systems Engineering. He received the Ph.D. degree in Electrical Engineering from Michigan State University, USA.
His current research interests include Internet of Things, detection of weak distribution patterns, network sampling and reconstruction, mining of network-based data for radicalization detection, and GAN based synthetic data generation. He is a member of Phi Kappa Phi, ACM and IEEE.
Contact him at anura.jayasumana@colostate.edu.
\end{IEEEbiography}
\end{document}